\documentclass[pra,twocolumn,showpacs,letterpaper,showpacs,superscriptaddress]{revtex4-1}
\usepackage{graphicx,amsmath,amssymb,amsfonts,latexsym,color,dcolumn,bm,subfigure}

\begin{document}

\title{Electrostatic patch effects in Casimir force experiments performed in the sphere-plane geometry}

\author{R. O. Behunin}
\affiliation{Center for Nonlinear Studies, Los Alamos National Laboratory, Los Alamos, New Mexico 87545, USA}
\affiliation{Theoretical Division, MS B213, Los Alamos National Laboratory, Los Alamos, New Mexico 87545, USA}
\author{Y. Zeng}
\affiliation{Theoretical Division, MS B213, Los Alamos National
Laboratory, Los Alamos, New Mexico 87545, USA}
\author{D. A. R. Dalvit}
\affiliation{Theoretical Division, MS B213, Los Alamos National
Laboratory, Los Alamos, New Mexico 87545, USA}
\author{S. Reynaud}
\affiliation{Laboratoire Kastler Brossel, CNRS, ENS, UPMC, Campus
Jussieu, F-75252 Paris France}

\date{ \today}

\begin{abstract}
Patch potentials arising from the polycrystalline structure of material samples may contribute significantly to measured signals in Casimir force experiments. Most of these experiments are performed in the sphere-plane geometry, yet, up to now all analysis of patch effects has been taken into account using the proximity force approximation which, in essence, treats the sphere as a plane. In this paper we present the exact solution for the electrostatic patch interaction energy in the sphere-plane geometry, and derive exact analytical formulas for the electrostatic patch force and minimizing potential. We perform
numerical simulations to analyze the distance dependence of the minimizing potential as a function of patch size, and quantify the sphere-plane patch force for a particular patch layout. 
Once the patch potentials on both
surfaces are measured by dedicated experiments our formulas can be used to exactly quantify the sphere-plane patch force in the particular experimental
situation.
 
\end{abstract}


\maketitle

\section{Introduction}

In distinction to what one is taught in introductory physics courses, the surfaces of real metals are not equipotentials but are rather described by a locally
varying surface voltage, known simply as patch potentials. Patch potentials exist for several reasons. One is that the work function of a crystalline structure depends upon which crystallographic plane an electron is extracted from \cite{Smoluchowski41,Gaillard06}. Real metal surfaces are typically composed of a network of crystallites with random crystallographic orientations, thereby giving rise to a nonuniform potential over the metal's surface. In addition, surface contamination by adsorbates is well-known experimentally \cite{Darling89,Camp91,Rossi92,Darling92} and theoretically \cite{Leung03} to lead to changes in the work function. Even for monocrystaline surfaces a spatially varying potential has been observed \cite{Pu10}. 
Patch potentials have important implications in various experimental disciplines, including gravitational measurements on elementary charged particles \cite{Fairbank67}, tests of the general theory of relativity \cite{Robertson06,Pollack08,Everitt11,Reasenberg11}, ion trapping \cite{Deslauriers06,Epstein07,Safavi11}, and the physics of Rydberg atoms \cite{Pu10,Carter11}. In this paper we focus on the effects that electrostatic patches can have on measurements of the Casimir force \cite{Speake03,Kim10,Kim10b,Lamoreaux10,LamoreauxLog,Behunin12}.

Most Casimir force measurements to date have been performed in the sphere-plane geometry 
in order to skirt alignment difficulties. The electrostatic interaction is 
used to calibrate the system and to determine the absolute separation between the sphere and the plane. In the idealized case of equipotential
surfaces, i.e., no patches on the surfaces, the exact analytical expression for the sphere-plane electrostatic force (the Coulomb force) is well known 
\cite{Smythe}, valid for arbitrary values of the ratio $D/R$, where $D$ is the sphere-plane separation and $R$ is the radius of the sphere. In typical experiments $D/R \ll 1$, and the exact expression reduces
to its proximity force approximation (PFA). This approximation replaces the sphere by infinitesimal planar surface elements and computes the
electrostatic force by adding plane-plane contributions, as if they were independent. For non-equipotential surfaces PFA has also been used
to compute the electrostatic patch force between the sphere and the plane \cite{Speake03,Kim10,LamoreauxLog,Behunin12}. A further assumption in the computation
of patch effects has been the ergodic hypothesis, that assumes that the actual realization of patches on both surfaces can be well represented by statistical properties of their sizes, shapes and voltages. Since the sphere has a compact cross section the sphere-plane interaction can be characterized by an effective area of interaction, and the ergodic hypothesis is expected to be satisfied when there are many patches within the interaction area, thereby providing a
fair representation of the patches' statistical properties.

The main goal of this paper is to derive the exact analytical expression for the sphere-plane electrostatic patch force which to the best of our knowledge has not yet been reported in the literature. 
Previous 
theoretical \cite{Kim10,Lamoreaux10,LamoreauxLog} and experimental \cite{Kim09,Sushkov11} works have used PFA to address
the implications of patch potentials on the electrostatic calibration process for Casimir force measurements. In this paper we also address the same
issue using our exact expression for the sphere-plane electrostatic patch force.

Measurements of the Casimir force between vacuum-separated bodies rely on an electrostatic calibration which is performed by sweeping through various values of  an externally applied potential between the bodies. This procedure generates a parabolic force curve as a function of applied voltage, the minimum of which identifies the  minimizing potential. In the absence of patches, the minimizing potential is independent of $D$, and an external voltage equal to the minimizing potential allows for the nullification of  all electrostatic forces. However, when patches are present this is no longer the case. As we will show below using our
exact expression for the sphere-plane patch force, the presence of patches on the samples implies that there is no external voltage that can nullify the electrostatic interaction - at most the force can be minimized by applying a voltage equal to the minimizing potential. Also, the existence of patches
implies that this minimizing potential is, in general, a function of sphere-plane distance $D$.  
However, we will show that when many patches are contained within the effective area of interaction the spatial variation of the minimizing potential is suppressed. In this regime the minimizing potential may appear, in experiment, to be distance independent. 
Even in this situation the residual electrostatic force is non-zero. 

With a detailed knowledge of the patch potentials on the sphere and plane surfaces, to be provided by dedicated ongoing and future measurements,  the results contained here can be used to exactly quantify the contribution of electrostatic patches to measured signals in Casimir force experiments. 


\section{Exact sphere-plane patch force}

We seek the solution to the boundary value problem for the electrostatic potential $V({\bf x})$ in the space between the sphere and the plane. The potential satisfies the Laplace equation
\begin{equation}
\nabla^2 V({\bf x}) = 0 ,
\end{equation}
subject to the following boundary conditions:
\begin{equation}
V({\bf x})|_{{\bf x} \in \textbf{P}} =  V_{\rm p}({\bf x})  \; , \;   V({\bf x})|_{{\bf x} \in \textbf{S}} =  V_{\rm s}({\bf x}).
\label{boundary-conditions}
\end{equation}
Here $V_{\rm p}({\bf x})$ and $V_{\rm s}({\bf x})$ are the potentials on the plane and the sphere, respectively, and ${\bf P}$ and ${\bf S}$ denote the set of points belonging to the plane and the sphere, respectively. 
Once we have the solution for the potential we can calculate electric field, ${\bf E} = - \nabla V$, and thus the electrostatic energy in the sphere-plane
configuration
\begin{equation}
{\rm E}_{\rm sp}  =  
\frac{\varepsilon_o}{2} \int_{\cal V} d^3 {\bf x} \;  (\nabla V)^2 ,
\end{equation}
where ${\cal V}$ denotes the volume between the sphere and the plane, and $\varepsilon_o$ is the permittivity of vacuum. Finally, the electrostatic sphere-plane force is obtained by taking minus the gradient of the energy, ${\bf F}_{\rm sp} = - \nabla{\rm E}_{\rm sp}$.

In the following subsections we will outline the techniques used to arrive at the exact solution for the potential given general electrostatic patchy boundary
conditions, the resulting interaction energy and force, and finally the minimizing potential. To this end we will make use of bispherical coordinates \cite{Morse}.
The key advantages of the bispherical coordinate system for this problem are that the Laplacian separates and that the two  surfaces on which we define our boundary conditions are described by $\eta$-constant surfaces. 


\begin{figure}
\begin{center}
\includegraphics[width=3.2in]{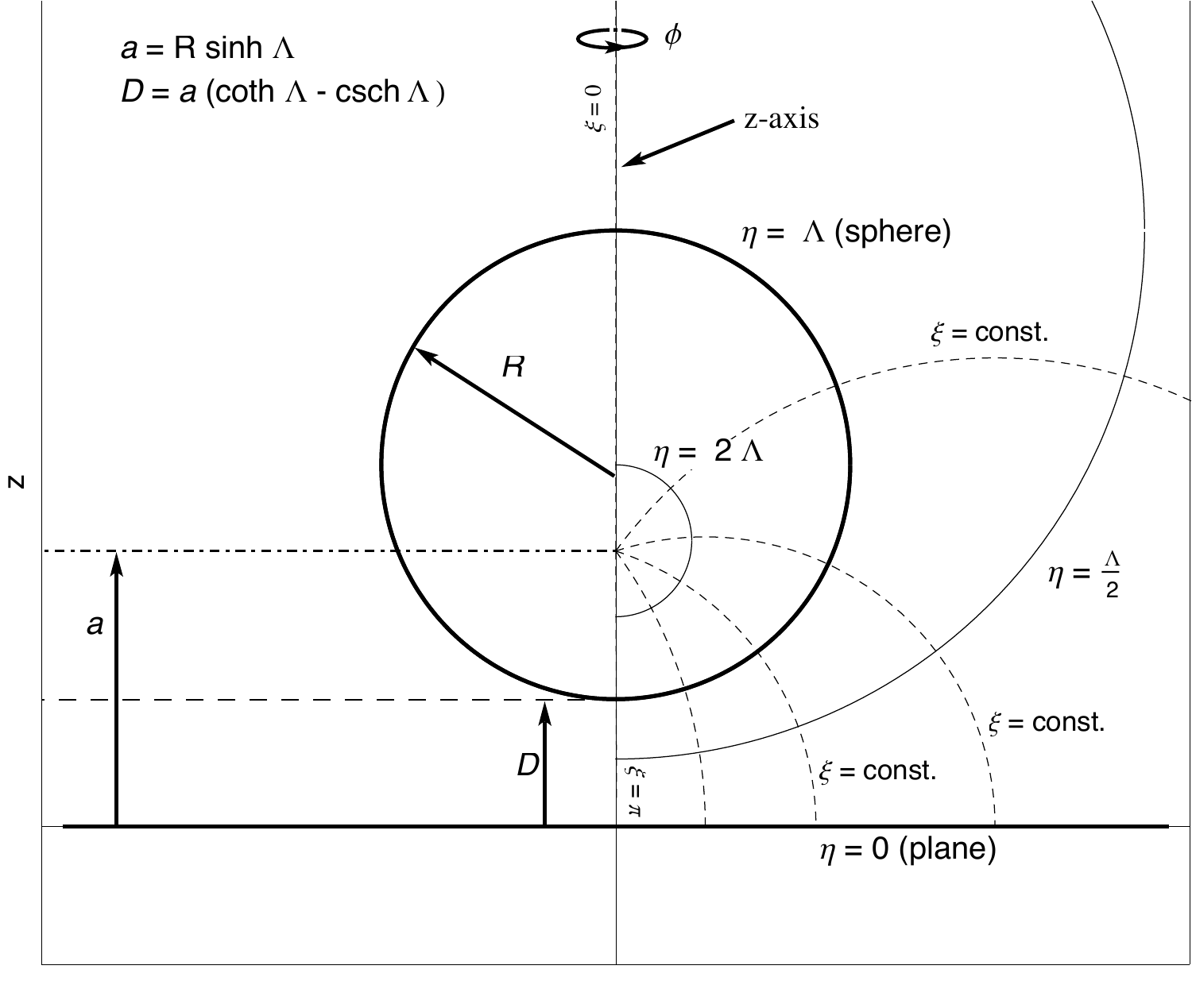}
\caption{Slice of the bispherical coordinate system along the $z$-axis. 
The solid lines are $\eta$-constant surfaces, and the dashed lines represent $\xi$-constant surfaces. 
To span $\mathbb{R}^3$ the coordinate system pictured above is rotated about the $z$-axis; the $\eta$-constant ($\eta \neq 0$) surfaces will sweep out spheres and the $\xi$-constant surfaces will sweep out ``apples" for $0<\xi<\pi/2$ and ``lemons" for $\pi/2< \xi <\pi$. The cross section of the $\eta = \Lambda$ surface 
corresponds to the sphere, and $\eta=0$ is the plane. The relations between the geometrical parameters $D$ and $R$ and the bispherical parameters
$a$ and $\Lambda$ are indicated in the figure.}
\label{bispherical-coords}
\end{center}
\end{figure}


\subsection{Electrostatic patch energy - Bispherical coordinates}

Bispherical coordinates $(\eta, \xi, \phi)$ can be used to label each point in $\mathbb{R}^3$. The correspondence with Cartesian coordinates is given by the following relations:
\begin{equation}
x = \frac{a \sin \xi \cos \phi}{\cosh \eta - \cos \xi}  \; ; \;  y = \frac{a \sin \xi \sin \phi}{\cosh \eta - \cos \xi}  \; ; \;   z = \frac{a \sinh \eta }{\cosh \eta - \cos \xi} .
\end{equation}
See Fig. \ref{bispherical-coords} for a visual representation.
For the sphere-plane geometry the adoption of bispherical coordinates leads to a significant simplification of the expression of the energy
through the use of Gauss' divergence theorem in curvilinear coordinates 
\begin{eqnarray}
\label{energy-general}
{\rm E}_{\rm sp} &= & 
\frac{\varepsilon_o}{2} \int_{\textbf{P}} d{\bf a} \cdot {\bf E} (\eta=0, \xi, \phi) V_{\rm p}(\xi, \phi) \\ \nonumber
& & -\frac{\varepsilon_o}{2} \int_{\textbf{S}}  d{\bf a} \cdot {\bf E} (\eta= \Lambda, \xi, \phi) V_{\rm s}(\xi, \phi) ,
\end{eqnarray}
where $E_\eta$ is the $\eta$ component of the electric field and $d{\bf a} = \sqrt{\Sigma(\eta)}d\xi d\phi \hat{\eta}$ is the oriented measure for the integration over an $\eta$-constant surface. The factor $\sqrt{\Sigma(\eta)} = (a^2 \sin \xi)/(\cosh \eta - \cos \xi)^2$ is the square root of the determinant of the metric induced on an $\eta$-constant surface. In these coordinates it is useful to relate $\Lambda$ (the $\eta$ coordinate for the sphere) and $a$ (the location of the foci of the bispherical coordinate system) to the radius of the sphere $R$ and the sphere-plane separation $D$:
\begin{eqnarray}
&& \cosh \Lambda = 1 + D/R , \nonumber \\
&& a = \sqrt{ (D+R)^2 - R^2} = R \sinh \Lambda.
\label{relations}
\end{eqnarray}

The Laplacian of the potential in these coordinates is
\begin{eqnarray}
\label{laplacian}
\nabla^2 V &=& \frac{(\cosh \eta - \cos \xi)^3}{ a^2 \sin \xi} \left[
\frac{\partial}{\partial \eta } \left( g(\eta, \xi) \frac{\partial V}{\partial \eta } \right) \right.
\nonumber \\
&& \left.
+ \frac{\partial}{\partial \xi } \left(  g(\eta, \xi)\frac{\partial V}{\partial \xi } \right)
+ \frac{g(\eta, \xi)}{\sin^2 \xi} \frac{\partial^2 V}{\partial \phi^2}  \right] ,
\end{eqnarray}
where $g(\eta, \xi) = \sin \xi /(\cosh \eta - \cos \xi)$. By making the ansatz $V = \sqrt{ \cosh \eta - \cos \xi} H(\eta) \Xi (\xi) \Psi(\phi)$ the Laplace equation separates, 
\begin{eqnarray}
&& \sin^2 \xi \frac{H''(\eta)}{H(\eta)} + \frac{ \Psi''(\phi)}{ \Psi(\phi)} - \frac{1}{4} \sin^2 \xi  \nonumber \\
&& + \sin \xi \cos \xi \frac{\Xi'(\xi)}{\Xi(\xi)} + \sin^2 \xi  \frac{\Xi''(\xi)}{\Xi(\xi)} = 0. 
\end{eqnarray}
To arrive at this equation we have divided through by $\csc^2(\xi) (\cosh \eta - \cos \xi)^{-5/2} H(\eta) \Xi(\xi) \Psi(\phi)$. 
Each of the functions $\Xi(\xi)$, $H(\eta)$, and $\Psi(\phi)$ can be solved for by separation of variables giving the 
general solution:
\begin{eqnarray}
&& V(\eta, \xi, \phi) = \sqrt{ \cosh \eta - \cos \xi}   \nonumber  \\
&& \times
\sum_{l = 0}^\infty \sum_{k = -l}^l e^{i k \phi} P^{k}_l (\cos \xi)
[ A_{l k}e^{\lambda_l \eta} + B_{l k}e^{-\lambda_l \eta} ] ,
\end{eqnarray}
where $\lambda_l = l + 1/2$, and $A_{kl}$ and $B_{kl}$ are constants to be determined by the boundary conditions on the potential. The function $P^k_l$ is 
the associated Legendre polynomial \cite{Gradshteyn}. 

The expansion coefficients are obtained by using the orthonormality properties of the eigenfunctions which serve as the basis for $V(\eta, \xi, \phi)$. By imposing the boundary conditions (\ref{boundary-conditions}), we find the exact solution for the electrostatic potential with general patchy boundary conditions
on the sphere and the plane, namely
\begin{widetext}

\begin{eqnarray}
\label{solution}
V(\eta, \xi, \phi) =   \sum_{\ell = 0}^\infty    \sum_{k = -\ell}^\ell
 \frac{\lambda_\ell}{ 2 \pi} 
(-1)^k
 \int_0^{2 \pi} d \phi'  \int_0^\pi d\xi' \sin \xi'
e^{i k (\phi - \phi')} P^{k}_\ell (\cos \xi) P^{-k}_\ell (\cos \xi') 
\nonumber \\
\times
\bigg[
\sqrt{ \frac{\cosh \eta - \cos \xi}{\cosh \Lambda -\cos \xi'}} \frac{\sinh \lambda_\ell \eta}{\sinh \lambda_\ell \Lambda} V_{\rm s}(\xi', \phi') -
\sqrt{\frac{\cosh \eta - \cos \xi}{1 -\cos \xi'}} \frac{\sinh \lambda_\ell( \eta-\Lambda)}{\sinh \lambda_\ell \Lambda} 
V_{\rm p}(\xi', \phi')  \bigg] .
\end{eqnarray}
Using (\ref{solution}) the general expression for the electrostatic energy of the sphere-plane system can be written by noting that the $\eta$-component of the electric field is  $E_\eta =  - a^{-1} (\cosh \eta -\cos \xi) \frac{\partial}{\partial \eta} V(\eta, \xi, \phi)$,
where we have used the expression for the gradient in curvilinear coordinates. By combining the result for $E_\eta$ with the expression for the energy (\ref{energy-general}) we find the general expression for the electrostatic energy
\begin{eqnarray}
&&{ \rm E}_{\rm sp} =  \frac{\varepsilon_o}{2} R \sinh \Lambda \sum_{\ell=0}^\infty \sum_{k = -\ell}^\ell \frac{\lambda_\ell}{2\pi}(-1)^k \int d\Omega \int d\Omega'  
e^{i k (\phi-\phi')} P^k_\ell(\cos \xi) P^{-k}_\ell(\cos \xi')   \nonumber \\
&&
\times \left[
\frac{V_{\rm s}(\Omega) V_{\rm s}(\Omega')}{ \sqrt{\cosh \Lambda -\cos \xi} \sqrt{\cosh \Lambda -\cos \xi'} }
\left( 
\lambda_\ell \coth \lambda_\ell \Lambda + \frac{\sinh \Lambda}{2(\cosh \Lambda - \cos \xi)} 
\right)  
+
\frac{V_{\rm p}(\Omega) V_{\rm p}(\Omega')}{ \sqrt{1 -\cos \xi} \sqrt{1-\cos \xi'} } \lambda_\ell  \coth \lambda_\ell \Lambda  \right. \nonumber \\
&&
\left.
- \frac{V_{\rm s}(\Omega) V_{\rm p}(\Omega')}{ \sqrt{\cosh \Lambda -\cos \xi} \sqrt{1 -\cos \xi'} } \frac{ \lambda_\ell}{\sinh \lambda_\ell \Lambda}   
- \frac{V_{\rm p}(\Omega) V_{\rm s}(\Omega')}{ \sqrt{1 -\cos \xi} \sqrt{\cosh \Lambda -\cos \xi'} } \frac{ \lambda_\ell}{\sinh \lambda_\ell \Lambda}   
\right] .
\label{exact-energy}
\end{eqnarray}

\end{widetext}
Here $ \int d\Omega$ is the integration over the ``solid angle'' $\int_0^\pi d\xi \int_0^{2\pi} d\phi \sin \xi$, and $V(\Omega) \equiv V(\xi, \phi)$ in bispherical coordinates.
 
Self-energy terms, both for the sphere and for the plane, are contained within Eq.(\ref{exact-energy}),
and their expression in bispherical coordinates can be found by taking the limit of infinite sphere-plane separation $D \rightarrow \infty$. We describe the calculation of these self-energy terms in Appendix A. 
In order to obtain the sphere-plane electrostatic interaction energy these self-energy terms must
be subtracted from the above expression for the energy.

 
\subsection{Electrostatic patch force}

We can find the sphere-plane patch force by taking the derivative of the interaction energy
with respect to the separation, 
${\rm F}_{\rm sp} = - (\partial/ \partial D) {\rm E}_{\rm sp}^{\rm int}$.
The expression for the energy in bispherical coordinates (\ref{exact-energy}) is, however, not very transparent for computing the force.
This is due to the fact that in bispherical coordinates  the sphere and plane potentials, which prescribe our boundary conditions, are a
function of the sphere-plane separation  \cite{explanation}.
To circumvent this complication and to connect with the natural basis in which the patch potentials are to be measured, we transform
to the natural coordinate system for each body. Therefore, we make the following change of variables (see Appendix A for details of coordinate
transformations from bispherical to spherical or polar coordinates)
\begin{eqnarray}
& & \int d \Omega \ V_{\rm s} (\Omega) (...) \to \int d \Omega_{\rm s} \ V_{\rm s} (\Omega_{\rm s}) \frac{\sinh^2 \Lambda}{(\cosh \Lambda + \cos \theta)^2} (...) , \nonumber \\
& &  \int d \Omega \ V_{\rm p} (\Omega) (...) \to \int d \Omega_{\rm p} \ V_{\rm p} (\Omega_{\rm p}) \frac{4 a^2}{(\rho^2+a^2)^2} (...) , 
\end{eqnarray}
where $\Omega_{\rm s} \equiv (\theta, \phi)$ are spherical coordinates on the sphere and $\Omega_{\rm p} \equiv (\rho, \phi)$ are polar coordinates on the plane. 
The integration measure $\int d \Omega_{\rm s} $ is given by $ \int_0^{2 \pi} d \phi \int_0^\pi d \theta \sin \theta$ with $\theta$ defined as the polar angle on the sphere,
and $\int d \Omega_{\rm p} = \int_0^{2 \pi} d \phi \int_0^\infty d \rho \ \rho$ where $\rho$ is the radius for a polar coordinate system defined on the plane. 
We can then express the sphere-plane electrostatic interaction energy in terms of the natural basis for the two bodies
\begin{equation}
{\rm E}_{\rm sp} = \sum_{a,b ={\rm s,p}} \int d \Omega_a \int d\Omega_b  V_a(\Omega_a) \mathcal{E}_{a,b} (\Omega_a; \Omega_b;D)  V_b(\Omega_b),
\label{energy-xfrm}
\end{equation}
where all dependence of the energy on the sphere-plane separation is now contained in the kernels $\mathcal{E}_{a,b} (\Omega_a; \Omega_b;D) $. 
Since the functions $\mathcal{E}_{a,b}(\Omega_a; \Omega_b;D)$ are complicated we will place their explicit derivations and expressions in Appendix B. 
By taking the derivative of the energy Eq.(\ref{energy-xfrm}) with respect to $D$, the exact electrostatic patch force between the sphere and the plane can now be computed
\begin{equation}
{\rm F}_{\rm sp} = \sum_{a,b =s,p} \int d \Omega_a \int d\Omega_b  V_a(\Omega_a) \mathcal{F}_{a,b} (\Omega_a; \Omega_b;D)  V_b(\Omega_b) ,
\label{exact-force}
\end{equation}
where $\mathcal{F}_{a,b} (\Omega_a; \Omega_b;D) = -(\partial/\partial D) \mathcal{E}_{a,b} (\Omega_a; \Omega_b;D)$.
The force above is general for arbitrary boundary conditions on the sphere and the plane. 

It is important to emphasize that in these expressions the origin of polar coordinate system
on the plane is assumed to be right below the sphere, at the point of closest approach between
the two bodies. Generally, the measured electrostatic patch potential distribution on the plane
will be done with respect to a different coordinate system, say a Cartesian system on the
plane. In this case, the appropriate change of coordinate system on the plane must be performed
prior to using Eqs.(\ref{exact-energy},\ref{exact-force}). 

In Appendix C
we show how to obtain from our exact expressions for the energy Eq.(\ref{exact-energy}) and force
Eq.(\ref{exact-force}) for patchy boundary conditions the corresponding well-known formulae for the
special case of equipotential surfaces.


\subsection{Minimizing potential and residual electrostatic force}

As mentioned above, in most Casimir force measurements an external voltage $V_{0}$ is applied between the two surfaces
to perform the calibration of the system. By sweeping $V_{0}$ between positive and negative values,
the total interaction force  (or its gradient) versus $V_{0}$ is measured for fixed sphere-plane
separation $D$, resulting in force vs. potential ``parabola" due to the quadratic dependence of 
${\rm F}_{\rm sp}$ on $V_{0}$.  These measurements are then repeated for each separation. The minima  of each of the parabolas defines the minimizing potential, 
namely
\begin{equation}
\label{minimizing-potential-definition}
\left. \frac{\partial {\rm F}_{\rm sp}}{\partial V_{0}} \right|_{V_{0}=V_{\rm min}} = 0.
\end{equation}
An explicit exact expression for the minimizing potential can be found in this way (see also 
\cite{Kim10,LamoreauxLog} for a similar approach using PFA).
To do so we replace in Eq.(\ref{exact-force}) the patchy potential on the sphere $V_{\rm s}$ by
$V_{0} +  V_{\rm s}(\Omega_{\rm s})$, i.e., by the addition of the constant applied potential and the non-constant
patchy one. 
Alternatively, we can do a similar replacement for the patchy potential on the plane and, of course, an identical
minimizing potential is obtained. 
The surface potentials on each of the objects will be assumed to fluctuate around the potential given by the 
average work function of the surface material.
For convenience we will write the potentials in terms of their average value and a term describing
fluctuations around zero. For example, the potential on the sphere becomes $V_{\rm s}(\Omega_s) = \bar{V}_{\rm s}+\Delta V_{\rm s}(\Omega_{\rm s})$ 
where $\bar{V}_{\rm s}$ denotes the average potential and $\Delta V_{\rm s} (\Omega_s)$ its fluctuations.  
By solving Eq.(\ref{minimizing-potential-definition}) we find
\begin{widetext}

\begin{align}
\label{minimizing-potential}
V_{\rm min} (D) = & - \frac{ 
\int d\Omega_{\rm s} \int d\Omega'_{\rm s}  (\bar{V}_{\rm s}+\Delta V_{\rm s}(\Omega_{\rm s}))\mathcal{F}_{{\rm s,s}}(\Omega_{\rm s};\Omega'_{\rm s};D) + 
\int d\Omega_{\rm p} \int d\Omega_{\rm s} (\bar{V}_{\rm p}+\Delta V_{\rm p}(\Omega_{\rm p}))  \mathcal{F}_{{\rm p,s}}(\Omega_{\rm p};\Omega_{\rm s};D)}
{\int d\Omega_{\rm s}  \int d\Omega'_{\rm s} 
\mathcal{F}_{{\rm s,s}} (\Omega_{\rm s};\Omega'_{\rm s};D)
} \nonumber 
\\
= & - \bar{V}_{\rm s} + \bar{V}_{\rm p} - \underbrace{ \frac{ 
\int d\Omega_{\rm s} \int d\Omega'_{\rm s}  \Delta V_{\rm s}(\Omega_{\rm s})\mathcal{F}_{{\rm s,s}}(\Omega_{\rm s};\Omega'_{\rm s};D) + 
\int d\Omega_{\rm p} \int d\Omega_{\rm s} \Delta V_{\rm p}(\Omega_{\rm p})  \mathcal{F}_{{\rm p,s}}(\Omega_{\rm p};\Omega_{\rm s};D)}
{\int d\Omega_{\rm s}  \int d\Omega'_{\rm s} 
\mathcal{F}_{{\rm s,s}} (\Omega_{\rm s};\Omega'_{\rm s};D)}
}_{\Delta V_{\rm min} (D)} .
\end{align}
In the second line above the integrations against the average potentials have been done. This result can be easily understood for the case 
when there are no patches altogether,  $\Delta V_{\rm s} = \Delta V_{\rm p} = 0$. In this case the minimizing potential is the applied potential necessary to nullify the force between an equipotential sphere and plane (see (\ref{eq-p})), and does not depend on $D$. The integrals against the fluctuating parts of the potential above,  $\Delta V_{\rm min}(D)$,  represent a sort of weighted average of the patchy part of the potential and are responsible for all of the spatial dependence of the minimizing potential. 

In general, the minimizing potential depends on the sphere-plane separation $D$ through the distance dependency of the kernels
$\mathcal{F}_{\rm s,s}$ and $\mathcal{F}_{\rm p,s}$. This will be shown explicitly in our numerical examples below, where we will also discuss the special
conditions under which $V_{\rm min}$ may appear to be distance-independent even in the presence of patches. 
One should also note that, in general,  setting the applied potential $V_0$ equal to the minimizing potential $V_{\rm min}$  does not nullify the electrostatic patch contribution to the total sphere-plane force. This can be seen by evaluating the electrostatic force at $V_0 = V_{\rm min}$:
\begin{align}
\label{minimized-force}
{\rm F}_{\rm sp}(V_{0}=V_{\rm min}) = & \sum_{a,b ={\rm s,p}} \int d \Omega_a \int d\Omega_b  V_a(\Omega_a) \mathcal{F}_{a,b} (\Omega_a; \Omega_b;D)  V_b(\Omega_b) -
V_{\rm min}^2 \int d\Omega_{\rm s} \int d\Omega'_{\rm s}  \mathcal{F}_{{\rm s,s}}(\Omega_{\rm s};\Omega'_{\rm s};D)  \nonumber \\
= & \sum_{a,b ={\rm s,p}} \int d \Omega_a \int d\Omega_b  \Delta V_a(\Omega_a) \mathcal{F}_{a,b} (\Omega_a; \Omega_b;D) \Delta V_b(\Omega_b) -
\Delta V_{\rm min}^2(D) \int d\Omega_{\rm s} \int d\Omega'_{\rm s}  \mathcal{F}_{{\rm s,s}}(\Omega_{\rm s};\Omega'_{\rm s};D)  ,
\end{align}
which is generally different from zero. The second line shows that the electrostatic calibration completely eliminates the equipotential component of the force,
but does not eliminate the fluctuating part. This residual electrostatic force, together with any other voltage-independent interactions (such as the Casimir force),
make up the signal in Casimir force measurements. 

The expression (\ref{minimized-force}) gives the minimum magnitude that the sphere-plane electrostatic force can take
for arbitrary surface potentials $\Delta V_{\rm a}$. Let us now consider what particular form $\Delta V_{\rm a}$ must take in order to minimize the residual force given by (\ref{minimized-force}). 
To do this we will take the variational derivative of ${\rm F}_{\rm sp}(V_{0}=V_{\rm min})$ with respect to the surface potentials and set the result to zero:
\begin{eqnarray}
0=\frac{\delta {\rm F}_{\rm sp}(V_{0} =  V_{\rm min}) }{ \delta \Delta V_a(\Omega_a)} &=& 2 \sum_{b =s,p}  \int d\Omega_b   \mathcal{F}_{a,b} (\Omega_a; \Omega_b;D)  \Delta V_b(\Omega_b) -
2 \Delta V_{\rm min}(D) \int d\Omega_s \int d\Omega'_s  \mathcal{F}_{ss}(\Omega_s;\Omega'_s;D)  \frac{\delta \Delta V_{\rm min}(D) }{ \delta \Delta V_a(\Omega_a)} 
\nonumber \\
&=& 2 \sum_{b =s,p}  \int d\Omega_b \bigg[  \mathcal{F}_{a,b} (\Omega_a; \Omega_b;D)   -
 \frac{ \int d\Omega_{\rm s} \int d\Omega'_{\rm s}  \mathcal{F}_{{\rm s},b}(\Omega_{\rm s};\Omega_b;D)  \mathcal{F}_{{\rm s},a}(\Omega'_{\rm s};\Omega_a;D)  }{ \int d \Omega_{\rm s} \int d\Omega_{\rm s}'   \mathcal{F}_{\rm s,s} (\Omega_{\rm s}; \Omega_{\rm s}';D)  } \bigg] \Delta V_b(\Omega_b) .
\end{eqnarray}
Since the above equation must be satisfied for all $D$ and the distance dependence cannot be factored out of the integral, the function in square brackets can take on nearly any value at each point in the integration domain. This leads to the solution for the above integral equation $\Delta V_a(\Omega_a) = 0$ for which it is easy to verify the ${\rm F}_{\rm sp}(V_{0}=V_{\rm min}) = 0$. Since this particular choice for the fluctuating potentials gives the extremum for the electrostatic force, any spatial variation of the potentials on the surfaces leads to a non-vanishing patch force. This proves that it is impossible to nullify the electrostatic force  by an externally applied potential when patches are present.

\end{widetext}


\subsection{Insights on the spatial dependence of the minimizing potential from the PFA}

Before discussing the results of our numerical simulations, we would like to give some theoretical arguments related to the conditions
under which the minimizing potential depends on distance, and to what one can infer about patches in the cases where the minimizing potential
is distance independent. To begin let us explain the reasons why an externally applied potential is necessary. 

In Casimir force experiments  there exists an intrinsic potential difference between the samples, the contact potential, $V_{\rm con}$, 
which is an average surface potential difference
whose 
physical origin is the electrical connections between the two bodies, differences in work function between the samples, and the presence of patches. 
In the following considerations we will neglect the effects from connecting wires, solder joints, etc., and focus entirely on the contact potential difference arising from work function differences and patches.
The contact potential leads to electrostatic forces between the bodies which can dominate over the desired Casimir force signal. 
By applying an appropriate bias voltage (the minimizing potential) the additional force arising from $V_{\rm con}$ can be nullified. 
Thus, we can understand the nature of the minimizing potential via its direct relationship to $V_{\rm con}$. 
We should emphasize that this bias voltage does not nullify the total electrostatic force, that has components arising from fluctuating patch voltages
which are not accounted for in $V_{\rm con}$.

Our goal now is to try to understand the spatial dependence of the contact potential. To do this we will lay out two ideas; the first quantifies the electrostatic sphere-plane force via an effective area of interaction, and the second relates the contact potential to a weighted average of patch voltages. To begin, consider two patchy surfaces interacting within the PFA limit, where the sphere is treated as a large but finite plane. In distinction to the case of two infinite planes the sphere-plane force can be characterized by an effective area of interaction. This is not surprising given that the sphere has a finite cross section. Within the PFA we can estimate this effective area of interaction by equating the sphere-plane force to the product of the plane-plane pressure and the effective area of interaction, ${\rm F}_{\rm sp} = {\rm P}_{\rm pp} A_{\rm eff}$,
where ${\rm F}_{\rm sp}$ is the sphere-plane force and ${\rm P}_{\rm pp}$ is the plane-plane electrostatic pressure. In the PFA limit the sphere-plane force can be approximated by the plane-plane energy per area ${\rm E}_{\rm pp}$,
${\rm F}_{\rm sp} \approx 2\pi R {\rm E}_{\rm pp}$. By noting that ${\rm E}_{\rm pp} \sim D {\rm P}_{\rm pp}$ we can solve for the effective area of interaction giving

\begin{equation}
\label{ }
A_{\rm eff} \sim 2 \pi R D.
\end{equation} 

Now we will connect the idea of an effective area of interaction with the contact potential.  To do so we will assume that each of the bodies are polycrystalline structures for which the work function varies above the surface due to the different local grain crystallographic orientation.  Given the variation of the potential over the surface we will roughly identify the contact potential with the average work function difference between the samples observed within the effective area. Thus, we can write the contact potential formally as
\begin{equation}
\label{contact-potential}
V_{\rm con} \sim \sum_{i,j}^N V_{ij} w_{ij},
\end{equation}
where the indices $i$ and $j$ are used to label the patches on the surfaces, $N$ is the number of patches within the effective area, and $w_{ij}$ is a normalized weight which is meant to roughly account for the fact that patches far from the point of the sphere and plane's nearest approach should contribute less to the value of the contact potential. 

In order to make these assertions more precise we will derive an expression for the minimizing potential (which equals the contact potential) within the PFA limit.
We should stress that this analysis will only roughly characterize the spatial variation of the minimizing potential and is used here to gain physical insight.

Our starting point is the expression for the sphere-plane force found in \cite{Kim10}:

\begin{equation}
\label{Fsp-PWS}
{\rm F}_{\rm sp} \approx \frac{\varepsilon_0}{2} \int_0^{2\pi} d \phi \int_0^R d \rho \ \rho  \frac{ (V_0 - V(\Omega_{\rm p}))^2}{(D+\rho^2/2R)^2} ,
\end{equation}
where $V(\Omega_{\rm p})$ is the spatially varying potential difference between the two plates,
measured at the the polar coordinate on the plane described by $\Omega_p$ (note that $V(\Omega_{\rm p})$ should not be
confused with the patch potential on the plane, $V_{\rm p}(\Omega_{\rm p})$).
By using Eq.(\ref{minimizing-potential-definition}) we can compute the 
minimizing potential (this equation also appears in \cite{Kim10} see Eq.(22)). After replacing 
$V(\Omega_{\rm p})$ with its average value and its fluctuating component, i.e., $V(\Omega_{\rm p}) = \bar{V} + \Delta V(\Omega_{\rm p})$, we find
\begin{equation}
\label{}
V_{\rm min}=  \bar{V} + \frac{ \int_0^{2\pi} d \phi \int_0^R d\rho \ \rho  \frac{ \Delta V(\Omega_{\rm p})}{(D+\rho^2/2R)^2} }{ \int_0^{2\pi} d \phi \int_0^R d\rho \ \rho  \frac{ 1 }{(D+\rho^2/2R)^2} }.
\end{equation}

\begin{figure}[t]
\begin{center}
\includegraphics[width=3.4in]{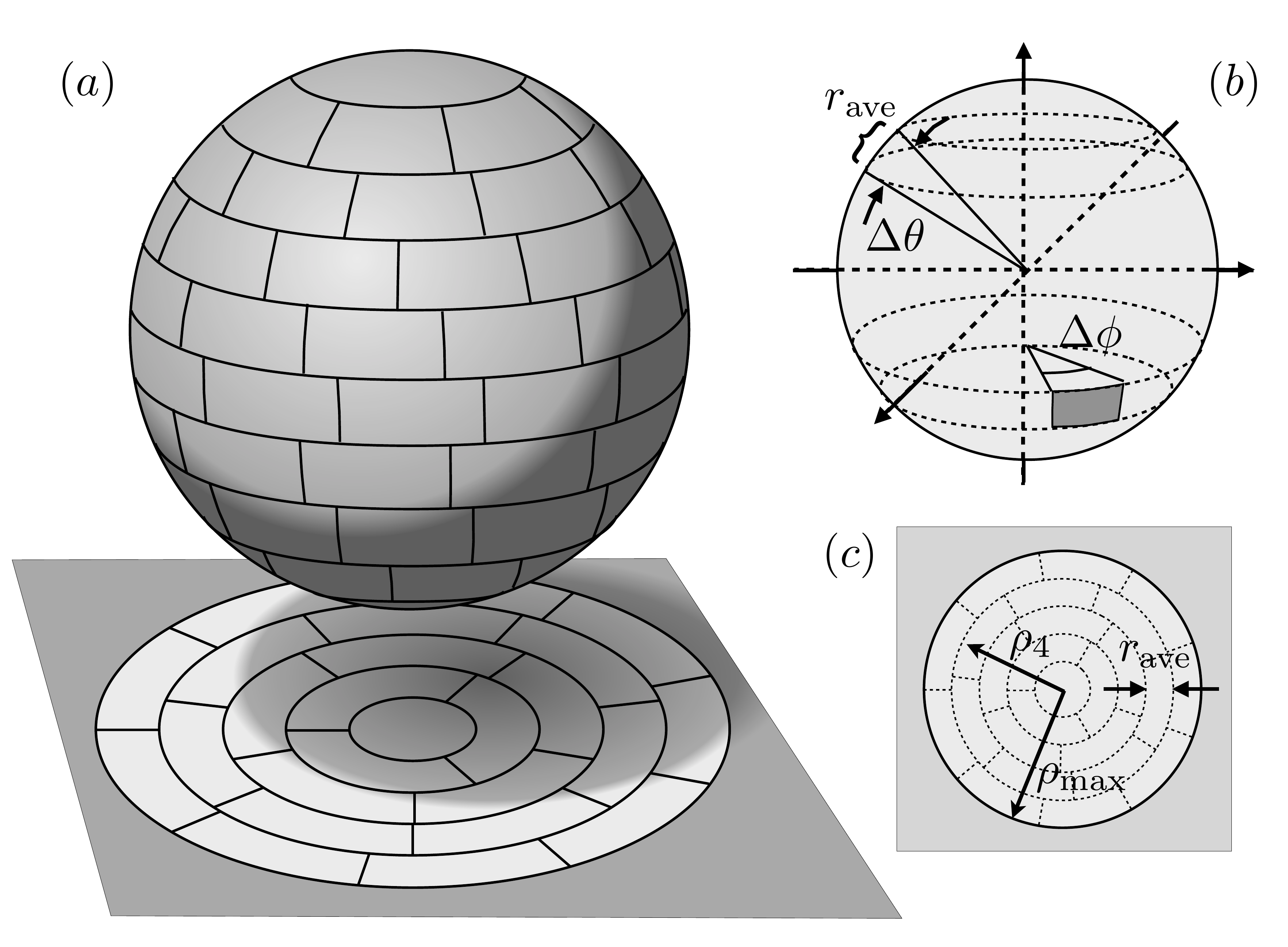}
\caption{(a) Illustration of the tesselation used for our simulations;
(b) Rings chosen on the sphere; (c) Patch layout
on the plane. In the numerics the parameter $\rho_{\rm max}$ is chosen to be larger than $5 R$
(although in (a) $\rho_{\rm max} \approx R$ for illustration purposes).
For the  PFA analysis of the spatial dependence of the minimizing potential both 
plates are tesselated as is depicted in (c).}
\label{TesselationDiagram}
\end{center}
\end{figure}

In order to proceed we will assume that $\Delta V(\Omega_{\rm p})$ is a piecewise constant function, and we will prescribe the geometry of the patches. The simplest choice for the patch layout is to break each surface into rings centered 
at the point of closest approach and then to divide each annulus into several pieces (see Fig. \ref{TesselationDiagram}c). Adopting this prescription, evaluating the integral in the denominator above, and using the fact that $D \ll R$, we can write the minimizing potential as 
\begin{equation}
\label{vmin-tesselated}
V_{\rm min} =  \bar{V} +  \sum_{i = 1}^M \sum_{j=1}^{K(i)}  \int_{\phi_{j-1(i)}}^{\phi_{j(i)}} \frac{d \phi}{2 \pi} \int_{\rho_{i-1}}^{\rho_{i}} \frac{d \rho}{R}  \frac{ \rho D\Delta V_{ij} }{(D+\rho^2/2R)^2} .
\end{equation}
Above, the indices $i$ and $j$ label each of the patches, $i$ denoting the ring and $j$ denoting the angular sector. The term $\Delta V_{ij}$ gives the fluctuating part of the potential which is constant within the boundary of each patch. 
The total number of rings is given by $M$, $K(i)$ is the number patches in the $i$th ring, the coordinates $\phi_j(i)$ denote the angular boundaries of the patch sectors in the $i$th ring, and $\rho_i$ denotes the outer radius of the $i$th ring.
For simplicity we will consider the case where $\rho_i = i r_{\rm ave}$ and divide the $i$th ring into $2i -1$ equal pieces. In this way each patch has a fixed area $\pi r_{\rm ave}^2$ and 
our expression above for the minimizing potential can be simplified. This discretization
of the surface certainly will not correspond precisely with a realistic distribution of 
patches on the sample surfaces and is not unique either. Despite this limitation we believe that this
crude approximation can provide some insights into the behavior of the minimizing
potential. 
 
After applying this discretization we find the simplified expression for the minimizing potential
\begin{equation}
\label{PFAmin}
V_{\rm min} =  \bar{V} + \sum_{i = 1}^M \sum_{j=1}^{2i-1}  \frac{  \frac{r_{\rm ave}^2}{ 2 D R} \Delta V_{ij}}{
\left( 1 + (i-1)^2 \frac{r_{\rm ave}^2}{ 2 D R} \right) \left(1+ i^2 \frac{r_{\rm ave}^2}{ 2 D R}\right)} ,
\end{equation}
where $M = R/r_{\rm ave}$ (since $M$ must be an integer one should take $M$ to be the floor of $R/r_{\rm ave}$ in numerical computations). 
Since the minimizing and contact potentials are equal we can use (\ref{PFAmin}) with (\ref{contact-potential}) to obtain an expression for the weights
\begin{equation}
\label{ }
w_{ij} =  \frac{  \frac{r_{\rm ave}^2}{ 2 D R} }{\left( 1 + (i-1)^2 \frac{r_{\rm ave}^2}{ 2 D R}\right) \left(1+ i^2 \frac{r_{\rm ave}^2}{ 2 D R}\right)}.
\end{equation}
It is interesting to note that these weights are parameterized by $r_{\rm ave}^2/2DR$, which is the ratio of the patch area to the effective area of interaction. One can see that when many patches fit inside the effective area of interaction, i.e., $r_{\rm ave}^2 / 2 D R \ll 1$, then $w_{ij} \propto r_{\rm ave}^2 / 2 D R$ for small $i$ (innermost rings), and becomes successively smaller for larger rings. At the outermost ring $w_{ij} \approx \frac{r_{\rm ave}^2}{ 2 D R} \frac{4 D^2}{R^2}$. Therefore, in this case, all rings contribute to the average but their influence is suppressed for large $i$ as $1/i^4$. 
In comparison, when the patch area is much larger than the effective area of interaction, i.e. $r_{\rm ave}^2 / 2 D R \gg 1$, the weight for the innermost ring is close to $1$, and the weights for all other rings are strongly suppressed, roughly proportional to $2 D R / r_{\rm ave}^2$. Therefore, in the large patch scenario only the patch located at the position of closest approach contributes to the contact potential. 

So far we have considered the contact potential for a fixed micro-realization of patches, meaning that the voltages and geometry of each patch have been assigned and fixed. This is the case one would encounter in an experiment. At this stage our calculation cannot proceed without a direct knowledge of the patch layout on the surfaces. However, based on the simple arguments made above it is not hard to make some qualitative statements about the expected sample-to-sample fluctuations of the contact potential as a function of separation. These arguments would apply to statistics on an ensemble of minimizing potential measurements performed with samples fabricated in the same way. 
Since the contact potential is roughly an average, the more patches which contribute to the average the more suppressed will be the sample-to-sample fluctuations.  This can be roughly understood because the uncertainty in an average value scales like $1/\sqrt{N}$ where $N$ is the number of data points used to compute the average. 
Therefore, a small patch size will lead to small minimizing potential fluctuations because more patches will determine the contact potential. The converse is true for large patches. For the same reasons large separations will lead to suppressed minimizing potential fluctuations since the effective area of interaction grows with separation. Likewise, at small separations as the ratio $r^2_{\rm ave}/2 DR$ becomes large one should expect large sample-to-sample fluctuations of the contact potential.
 
We can make these qualitative assertions more concrete by making statistical assumptions about the patch voltages. If each of the potentials is assumed to be assigned randomly and statistically independently of one another, and with the same variance, then the expected variation of the sum on $j$ takes the form
$\sum_{j=1}^{2i-1} \Delta V_{ij} = \pm \sqrt{2i-1} V_{\rm rms}$, 
where $V_{\rm rms}$ characterizes the expected rms fluctuations of the assigned potential for a single patch. 
This leads to the expected range of minimizing potential values given by

 \begin{equation}
\label{vmin-stdev}
  \bar{V} \pm \underbrace{ V_{\rm rms} \sum_{i = 1}^M  \frac{  \frac{r_{\rm ave}^2}{ 2 D R} \sqrt{2i -1}}{( 1 + (i-1)^2 \frac{r_{\rm ave}^2}{ 2 D R})(1+ i^2 \frac{r_{\rm ave}^2}{ 2 D R})}}_{\sigma_{V_{\rm min}}(D)}.
\end{equation}
Thus, in a given experiment, i.e. one micro-realization of patches, we would expect the minimizing potential to vary in position within  a few $\sigma_{V_{\rm min}}(D)$ of $\bar{V}$.
The variation is suppressed at large distances because averaging is performed over larger and larger effective areas.
However in the case of large patches, $r_{\rm ave}^2 / 2 D R \gg 1$, the $i=1$ term in the sum dominates and the expected variation is given roughly by 
$V_{\rm rms} $. 

Above we have made some rough arguments in order to characterize the expected variation of the contact potential. We should stress that this expected variation will tell us nothing about the minimizing potential in a single experiment: equation (\ref{vmin-stdev}) will not yield a prediction for the minimizing potential as a function of distance. Rather, the arguments above apply to an ensemble of measurements of the minimizing potential for different samples prepared in the same way. The expected fluctuations will only tell us the envelope within which roughly $66\%$ of minimizing potential measurements of various samples will lie.

The above analysis makes it clear that, in principle, the minimizing potential always depends on distance when patches are present. However, when the expected variation of the contact potential is much smaller than the uncertainty in the measurements, the minimizing potential will appear in practice to be independent of distance. This behavior is expected when the typical patch size is much smaller than the effective area of interaction, and as we will soon show, can be accompanied by a non-vanishing patch force.


\section{Numerical results}

In this section we will present our numerical results for the sphere-plane minimizing potential (\ref{minimizing-potential}) and residual electrostatic force 
(\ref{minimized-force}) for various patch sizes.


To compute the minimizing potential we evaluate Eq.(\ref{minimizing-potential}) in several steps. First, the sphere and the plane are tesselated (see Fig. \ref{TesselationDiagram}), and then each patch is independently and randomly assigned a potential from the values $(5.15, 5.04, 5.10) {\rm V}$, corresponding to the
work functions of the three principal crystallographic orientations of gold \cite{Singh09}.  For each micro-realization, defined by 
a given layout of potentials and patch geometries, we compute the minimizing potential using  (\ref{minimizing-potential}).
Given an ensemble of minimizing potential ``measurements" we perform statistics to recover the mean value and the expected variation. 
For numerical simplicity we keep the geometry of the patch layout fixed in all micro-realizations. 
Another point to acknowledge is that we assume that the average potential for the sphere and the plane are equal, as if they are made from the same material, and therefore $\bar{V}_{\rm p} = \bar{V}_{\rm s}$.

The tesselation is performed using a ring-like division of the sphere and the plane as is depicted in Fig. \ref{TesselationDiagram}a. 
The plane is divided into annuli in the same manner described above for the PFA treatment of the spatial dependence of the minimizing potential. The outer radius of the $i$th annulus, $\rho_i$, is given by $i r_{\rm ave}$ and therefore the width of each annulus is given by $r_{\rm ave}$. The $i$th ring is divided into $2i-1$ equal pieces so that  all patches have the same area. In order to manage the computation time we set the patch potentials to the average potential for radii greater than $\rho_{\rm max}$, which is chosen large enough so that all results converge.  As we mentioned in our PFA analysis above,  we
acknowledge
that this type of tesselation of the plane does not correspond with any realistic layout of patches on the plane.  We believe, however, that this approximation is sufficient to gain some insight about the patch size dependence of the force. Next, the sphere is tesselated in a similar manner by being divided into a fixed number of latitudinal lines any neighboring pair subtending polar angle $\Delta \theta$, afterward the lines of longitude dividing a given ring are chosen so that all of the patches have the same area. We choose $\Delta \theta$ to take the value $R \Delta \theta = r_{\rm ave}$. For this particular tesselation the polar angle describing the upper boundary of the $i$th ring is given by $\theta_i = \pi - \frac{r_{\rm ave}}{R} i$ where we have chosen $i=0$ to correspond with the polar angle of the ``south pole". With the particular choice we have adopted for the polar angles of the rings  we find the number of patches in the $i$th ring on the sphere is given by $ |{\rm Floor}[ (2 R^2/r_{\rm ave}^2) (\cos \theta_{i-1} - \cos \theta_{i})]|$.

\begin{figure}[t]
\begin{center}
 \includegraphics[width=2.5in]{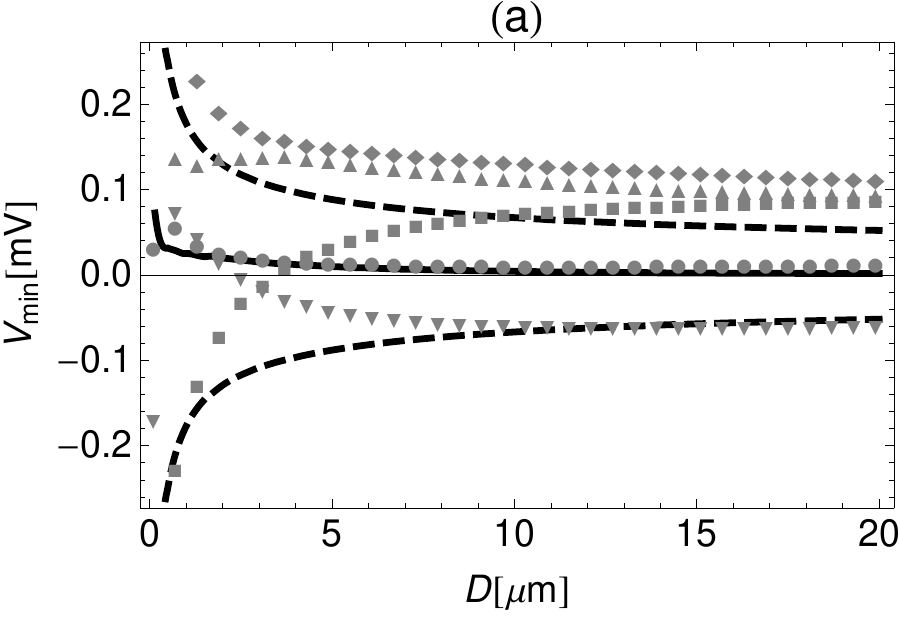} 
\\
 \includegraphics[width=2.5in]{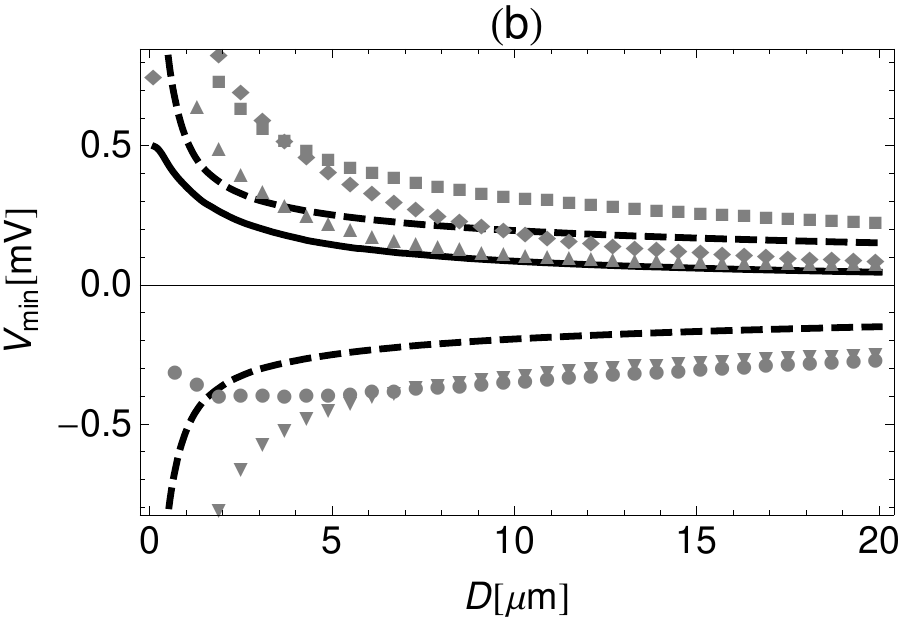} 
 \\
 \includegraphics[width=2.5in]{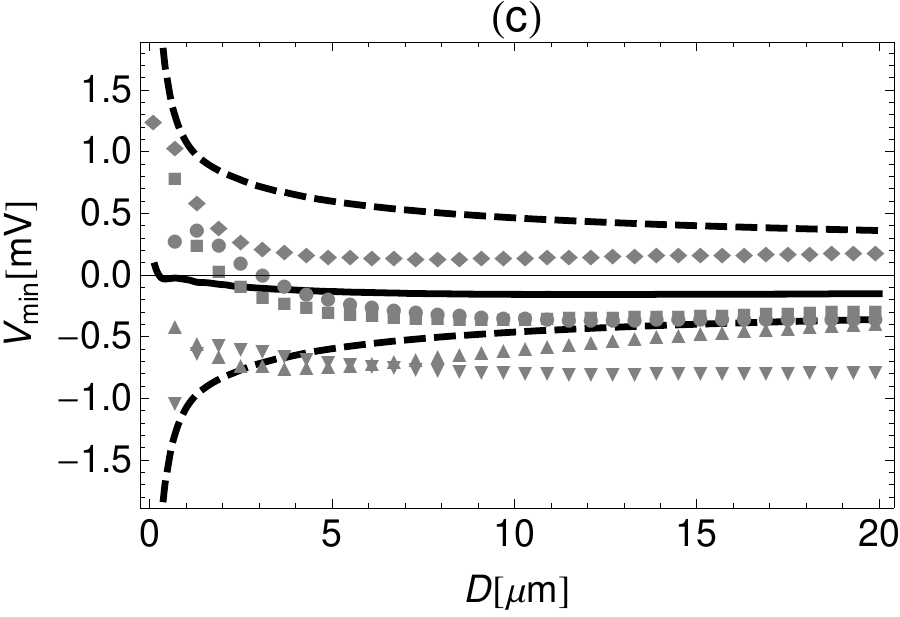} 
\end{center}
\caption{Minimizing potential as a function of distance for various patch sizes: (a) 100 nm, (b) 300 nm, and (c) 600 nm.
The solid curve 
is the average minimizing potential computed from 400 micro-realizations. The dashed curves
enclose the range of expected fluctuations computed from the standard deviation of all minimizing 
potential values at a fixed distance. The data points denote the minimizing potential
for five random micro-realizations. The radius of the sphere is $R=150 \mu$m.
}
\label{vmin-plot}
\end{figure}

The numerical results for the minimizing potential are presented in Fig. \ref{vmin-plot}. All of the plots employ the same 
convention: the solid line denotes the average minimizing potential as a function of distance, and the dashed lines are computed from
the standard deviation of the all minimizing potential values at each separation. Thus, by definition, the envelope created by the two
dashed lines contains roughly $66\%$ of all minimizing potential measurements. The data points indicated by the various plot markers are the values of the minimizing potential from five random micro-realizations. The figure illustrates the spatial dependence of the minimizing potential on patch size.
In line with our PFA analysis, as the patch size grows (from (a) to (c)) the expected minimizing potential fluctuations are enhanced. Additionally, for each patch size the
effective area of interaction decreases as the sphere-plane distance becomes smaller. Thus, the shorter the distance, the fewer the patches which
contribute to the average that determines the minimizing potential, and hence the larger are the fluctuations.

\begin{figure}[t]
\begin{center}
\includegraphics[width=3in]{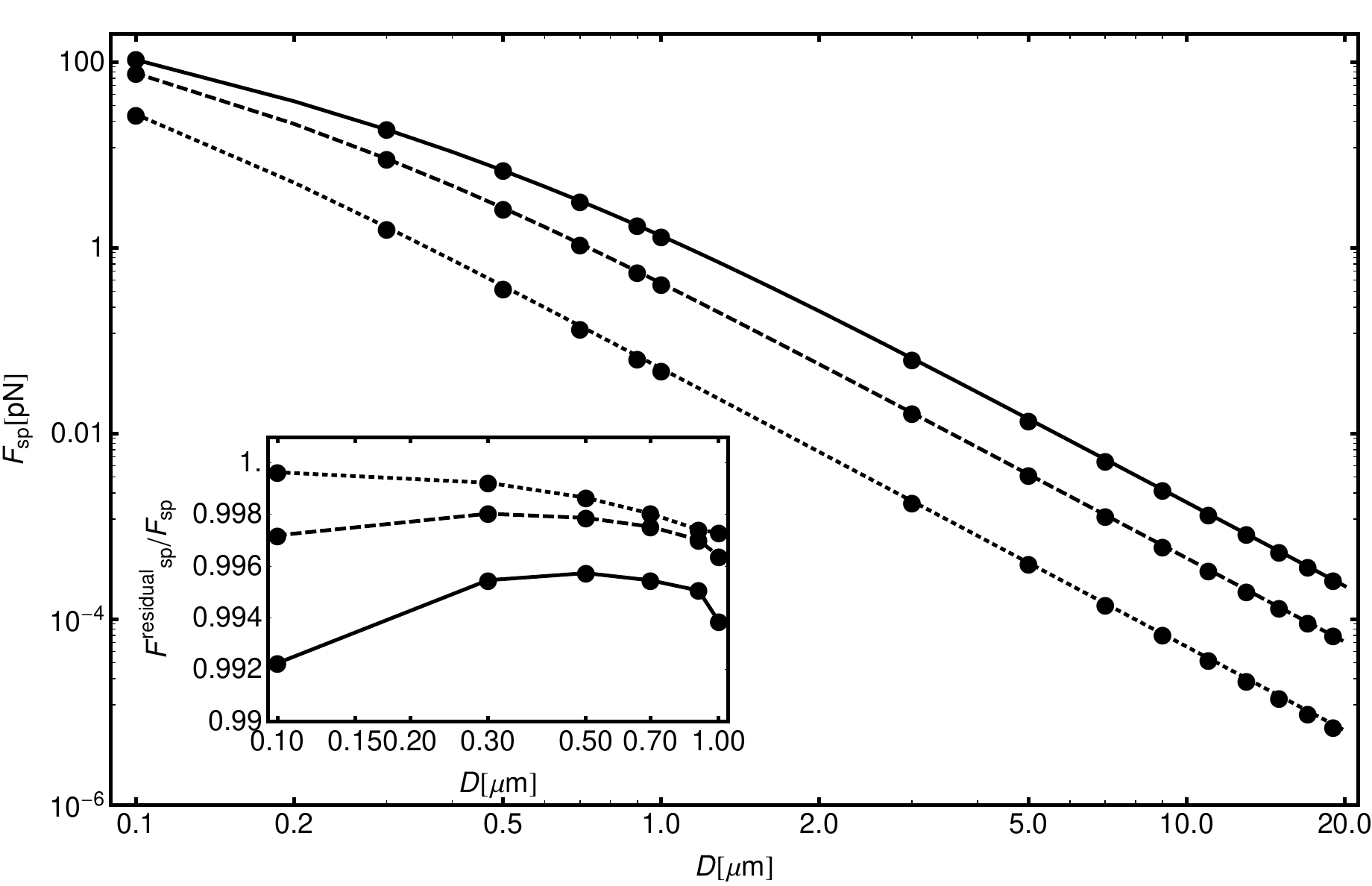}
\caption{Sphere-plane electrostatic patch force (\ref{exact-force}) as a function of separation for different patch sizes: 100 nm (dotted), 300 nm (dashed), and 600 nm (solid). The data points are the average patch force, the error bars are not visible at scale shown, and the lines are the patch force computed
with the voltage correlation method ($V_{\rm rms} = 45$mV). 
Inset: Ratio of the residual force (\ref{minimized-force}) to the
patch force (\ref{exact-force}) as a function of separation for different patch sizes.
The radius of the sphere is $R=150 \mu$m.}
\label{patchforce}
\end{center}
\end{figure}

To obtain the sphere-plane electrostatic patch force Eq.(\ref{exact-force}), we employ the tesselation described above, see Fig. \ref{TesselationDiagram}. Once we have an ensemble of realizations we find the average patch force and its expected fluctuations. The results of our simulations are depicted in  Fig. \ref{patchforce}. We also compare these numerical results  with 
the patch force computed by performing an average over voltage micro-realizations analytically while keeping the patch geometrical layout fixed.
To do this we compute the two-point voltage correlation functions on the plates, where we assume that the voltages on each patch are statistically independent, random variables (so-called ``quasi-local correlation", see Eqs. (10-12) of \cite{Behunin12}),
\begin{equation}
\langle V_a({\bf x}) V_b({\bf x}') \rangle_v = \delta_{ab} V_{\rm rms}^2 \sum_i \Theta_i({\bf x}) \Theta_i({\bf x}') ,
\label{correlation}
\end{equation}
where the function $\Theta_i({\bf x})$ is 1 for points ${\bf x}$ within the $i$th patch and 0 otherwise. Using this equation in (\ref{exact-force}) we
compute the ensemble averaged patch force $\langle F_{\rm sp} \rangle_v$. In Fig. (\ref{patchforce}) we compare the exact
numerics with the patch force computed via the voltage correlation method. As seen in the figure, the agreement between the two is excellent. 
In the inset we compare the residual electrostatic force (\ref{minimized-force}) with the above patch force (\ref{exact-force}). The two forces
are very similar, illustrating that the electrostatic calibration does not nullify the patch force.

\section{Conclusions}

In this paper we have derived the exact expression for the sphere-plane electrostatic potential, interaction energy and force for patchy boundary
conditions.  With knowledge of the potentials on the sample surfaces used in Casimir force experiments in the sphere-plane geometry, 
to be measured by dedicated experiments, these expressions can be used to exactly quantify the sphere-plane force due to patch potentials. 

As an added benefit of the exact solution we have been able to derive an exact formula for the minimizing potential used to calibrate many Casimir force experiments. We have found that when patches are present the minimizing potential always depends on position. However, the magnitude of the spatial fluctuations are controlled by the ratio of the typical patch size to the effective area of interaction. In the limit where this ratio is small the expected spatial modulation of the minimizing potential is suppressed, and provided that the expected variation is smaller than the experimental uncertainty it will appear to be independent of position. We have verified this behavior numerically.
Using proximity force arguments and numerical computations we were also able to show the following qualitative behavior: the typical fluctuations of the minimizing potential decay as a function of sphere-plane separation and correlate with the typical patch size. Our PFA analysis suggests that both of these observations can be understood in terms of a weighted average over the patches on both surfaces. When many patches contribute to the average (by having comparable weights) the expected fluctuations of the minimizing potential are suppressed.  This happens when many patches fit within the effective area of interaction.  In contrast, the expected fluctuations are greatly enhanced when the typical patch area is comparable to the effective area of interaction, as the patch at the point of nearest approach dominates the weighted average. 

Furthermore, we have derived an explicit formula for the residual electrostatic force (\ref{minimized-force})
that persists in Casimir force measurements after an electrostatic calibration has been performed.
By analyzing the residual electrostatic force we proved that the electrostatic interaction cannot be nullified by an applied field unless there are no patches. This behavior was also verified in our simulations. 


\section{Acknowledgments}

We would like to acknowledge fruitful discussions with Ricardo Decca, Francesco Intravaia, Woo-Joong Kim, Astrid Lambrecht, Steve Lamoreaux, Paulo Maia Neto, Alex Sushkov, and Andrew Sykes. 


\appendix
\section{Electrostatic self-energies}

In this Appendix we compute the large separation limit of the electrostatic sphere-plane energy (\ref{exact-energy}) in order to identify the self-energies
of the sphere and the plane in bispherical coordinates and show that they correspond to the well-known self-energies of each of these objects.

We start with the computation of the self-energy for the sphere. First, we briefly recall the derivation of its electrostatic self-energy in spherical coordinates. The electrostatic potential for an isolated sphere of radius $R$ with surface patch potentials $V_{\rm s}(\Omega) = V_{\rm s}(\theta,\phi)$ is given by the known solution of the Laplace equation in spherical coordinates,
\begin{eqnarray}
&& V({\bf x}) =\sum_{\ell = 0}^\infty    \sum_{m = -\ell}^\ell  \frac{\lambda_{\ell}}{2 \pi} (-1)^m \left( \frac{R}{r} \right)^{\ell+1}  \nonumber \\
&& \times \int d\Omega_{\rm s}' e^{i m (\phi-\phi')} P^m_{\ell}(\cos \theta)   P^{-m}_{\ell}(\cos \theta') V_{\rm s}(\Omega_{\rm s}') .
\end{eqnarray} 
Similar to the procedure used in the main text, we use the Gauss divergence theorem to express the sphere self-energy as
\begin{equation}
{\rm E}_{\rm s}^{\rm self} = - \frac{\varepsilon_o}{2} \int_S d{\bf a} \cdot V_{\rm s}(\Omega)
 \hat{\bf r}  \left. \frac{\partial}{\partial r} \;   V({\bf x}) \right|_{r=R} ,
\end{equation}
where the integration is over the sphere's surface. In this way we obtain the known electrostatic self-energy of the sphere in spherical
coordinates
\begin{eqnarray}
&& {\rm E}_{\rm s}^{\rm self} =   \frac{\varepsilon_o}{2} R  \sum_{\ell = 0}^\infty    \sum_{m = -\ell}^\ell  \frac{\lambda_{\ell} (\lambda_{\ell}+1/2)}{2 \pi} (-1)^m 
\int d\Omega_{\rm s} \int d\Omega'_{\rm s} \nonumber \\
&& \times e^{ i m (\phi-\phi')} 
P^m_{\ell}(\cos \theta)   P^{-m}_{\ell}(\cos \theta') V_{\rm s}(\Omega_{\rm s}) V_{\rm s}(\Omega'_{\rm s}) ,
\label{self-energy-sphere-spherical-coordinates}
\end{eqnarray}
where we used that $\lambda_{\ell}=\ell+1/2$. 

Next we show that by taking the large distance limit in (\ref{exact-energy}), the piece which is a quadratic function of the sphere's potential
 gives exactly this result for the sphere's self-energy. When $D \rightarrow \infty$, or equivalently when
$\Lambda \rightarrow \infty$, we get from  (\ref{exact-energy})
\begin{eqnarray}
&&{\rm E}_{\rm s}^{\rm self} = \lim_{\Lambda \rightarrow \infty} 
\frac{\varepsilon_o}{2} R  \sum_{\ell=0}^\infty \sum_{k = -\ell}^\ell \frac{\lambda_\ell (\lambda_\ell + 1/2)}{2\pi}(-1)^k \int d\Omega \int d\Omega'  \nonumber \\
&&
\times e^{i k (\phi-\phi')} P^k_\ell(\cos \xi) P^{-k}_\ell(\cos \xi') 
V_{\rm s}(\Omega) V_{\rm s}(\Omega') .
\label{self-energy-sphere}
\end{eqnarray}
We use the following identities that relate bispherical coordinates with spherical coordinates centered on the sphere
\begin{eqnarray}
\cos \xi  &=& \frac{1+\cosh \Lambda \cos \theta}{\cosh \Lambda + \cos \theta} ,  \\
\sin \xi &=& \frac{\sinh \Lambda \sin \theta}{\cosh \Lambda + \cos \theta} , \\
\sin \xi \ d \xi  &=& \frac{ \sinh^2 \Lambda}{(\cosh \Lambda + \cos \theta)^2} \sin \theta \ d \theta, \\
\frac{1}{(\cosh \Lambda -\cos \xi)^n } &=& \frac{(\cosh \Lambda + \cos \theta)^n }{\sinh^{2n} \Lambda} .
\end{eqnarray}
For $\Lambda\rightarrow \infty$ we obtain $\cos \xi \rightarrow \cos \theta$, $d \Omega \rightarrow d \Omega_{\rm s}$, and therefore
the sphere self-energy in bispherical coordinates (\ref{self-energy-sphere}) is identical to that in spherical coordinates 
(\ref{self-energy-sphere-spherical-coordinates}), as expected.

We now consider the self-energy for the plane. The electrostatic potential for an isolated plane located at $z=0$ having surface patch potentials $V_{\rm p}(\Omega) = V_{\rm p}({\bf r})$ (${\bf r}=(x,y)$)
is
\begin{equation}
V({\bf x})= \int \frac{d^2 {\bf k}}{(2 \pi)^2} e^{i {\bf k} \cdot {\bf r}} e^{-k z} V_{\rm p}({\bf k}) ,
\end{equation}
where $V_{\rm p}({\bf k})$ is the 2D Fourier transform of the patchy potential. The self-energy is
\begin{eqnarray}
{\rm E}_{\rm p}^{\rm self} &=&  \frac{\varepsilon_o}{2} \int d^2{\bf x} V_{\rm p}({\bf x}) \left. \frac{\partial}{\partial z} V({\bf x}) \right|_{z=0} 
\nonumber \\
&=& \frac{\varepsilon_o}{2} \int d^2{\bf r}  \int d^2{\bf r}' V_{\rm p}({\bf r}) f( {\bf r} - {\bf r}' ) V_{\rm p}({\bf r}') ,
\label{self-energy-plane-usual}
\end{eqnarray} 
where $f({\bf r}- {\bf r}') = (2 \pi)^{-2} \int d^2{\bf k} |{\bf k}| e^{i {\bf k} \cdot ({\bf r}-{\bf r}')}= - (2 \pi)^{-1} | {\bf r}-{\bf r}'|^{-3}$. 

As for the case of the self-energy for the sphere, we take the $D \rightarrow \infty$ limit (or, equivalently, the $a \rightarrow \infty$ limit) 
of (\ref{exact-energy}) and focus on the
term that depends on the plane. We get
\begin{eqnarray}
&&{\rm E}_{\rm p}^{\rm self} = \frac{\varepsilon_o}{2} a \sum_{\ell=0}^\infty \sum_{k = -\ell}^\ell \frac{\lambda_\ell^2}{2\pi}(-1)^k 
\int  d\Omega \int d\Omega'
\nonumber \\
&&
\times \frac{ 
e^{i k (\phi-\phi')} 
P^k_\ell(\cos \xi) P^{-k}_\ell(\cos \xi') 
V_{\rm p}(\Omega) V_{\rm p}(\Omega') }
{\sqrt{1-\cos \xi} \sqrt{1-\cos \xi'}}.
\label{self-energy-plane}
\end{eqnarray}
In order to show that this is identical to (\ref{self-energy-plane-usual}) we first
perform the summation over $k$ in Eq.(\ref{self-energy-plane}). For this we use the addition
theorem of Legendre polynomials,
$\sum_{k=-\ell}^{\ell} (-1)^k e^{i k (\phi-\phi')} 
P^k_\ell(\cos \xi) P^{-k}_\ell(\cos \xi')  = P_{\ell}(\cos \gamma)$,
where 
\begin{equation}
\cos \gamma = \cos \xi \cos \xi' + \sin \xi \sin \xi' \cos(\phi-\phi') .
\label{cosgamma}
\end{equation}
In the resulting equation we then change coordinate system from bispherical coordinates to polar coordinates on the plane ($\eta=0$), centered at the point of closest approach between the sphere and the plane, i.e. right below the sphere. The polar coordinate $\rho$ relates to the bispherical coordinate $\xi$ through
$ \rho = a \cot \xi/2$.  The following identities are useful to relate both coordinate systems:
\begin{eqnarray}
\cos \xi &=& \frac{(\rho/a)^2-1}{ (\rho/a)^2 +1 } , 
\label{cos-bispherical-polar}
\\
\sin \xi &=& \frac{ 2(\rho/a)}{ (\rho/a)^2 +1 } , 
\label{sin-bispherical-polar}
\\
\frac{\sin \xi d\xi}{ \sqrt{1-\cos \xi}}  &=& -  \frac{2 \sqrt{2} \rho d\rho }{a^2 [1+(\rho/a)^2]^{3/2} } .
\end{eqnarray}
Then Eq.(\ref{self-energy-plane}) takes the form
\begin{align}
{\rm E}_{\rm p}^{\rm self} = \lim_{a \rightarrow \infty}
\frac{2 \varepsilon_o}{\pi a^3} \int d\Omega_{\rm p} \int d\Omega_{\rm p}'
V_{\rm p}(\Omega_{\rm p}) V_{\rm p}(\Omega_{\rm p}') \nonumber \\
\times 
\sum_{\ell=0}^{\infty} \lambda_{\ell}^2 P_{\ell}(\cos \gamma) , 
\label{self-energy-plane-intermediate}
\end{align}
where $\Omega_{\rm p} = (\rho, \phi)$ are the polar coordinates on the plane, and
$d \Omega_{\rm p}$ is the corresponding measure. The above summation can be evaluated
using the generating function of Legendre polynomials, that verifies
\begin{equation}
S(t,u) \equiv \sum_{\ell=0}^{\infty} t^{\ell} P_{\ell}(u) = \frac{1}{\sqrt{1-2 t u + t^2}} .
\end{equation}
Recalling that $\lambda_{\ell}=\ell+1/2$, we obtain
\begin{eqnarray}
\sum_{\ell=0}^{\infty} \lambda^2_{\ell} t^{\ell} P_{\ell}(u) &=& 
\left( t \frac{\partial}{\partial t} t \frac{\partial}{\partial t} + t \frac{\partial}{\partial t}  + 
\frac{1}{4} \right) S(t,u)  \nonumber \\
&=& \frac{1-10 t^2 + t^4 + 4 t (1+t^2) u}{4 (1+t^2 - 2 t u )^{5/2}} \nonumber \\
&\equiv & \mathcal{G}(t,u) .
\label{identity-1}
\end{eqnarray}
Setting $t=1$ we derive the identity
$\sum_{\ell =0}^{\infty} \lambda_{\ell}^2 P_{\ell}(\cos \gamma) = - [2 ( 1- \cos \gamma)]^{-3/2}$.
Using (\ref{cos-bispherical-polar}) and (\ref{sin-bispherical-polar}) in the definition of
$\cos \gamma$ and taking the $a \rightarrow \infty$ limit we obtain
$1- \cos \gamma \approx 2 |{\bf r}-{\bf r}'|/a^2$. Hence (\ref{self-energy-plane-intermediate}) is equal to
\begin{equation}
{\rm E}_{\rm p}^{\rm self} = 
- \frac{\varepsilon_o}{4 \pi} \int d^2{\bf r} \int d^2{\bf r}' V_{\rm p}({\bf r}) 
|{\bf r}-{\bf r}'|^{-3} V_{\rm p}({\bf r}') ,
\end{equation}
which is exactly the plane self-energy in polar coordinates (\ref{self-energy-plane-usual}), as it should.


\section{Derivation and expression of sphere-plane energy kernels}

In this Appendix we give the derivation of the kernels appearing in Eq.(\ref{energy-xfrm}).
We first write the energy (\ref{exact-energy}) in a compact
form, ${\rm E}_{\rm sp} = {\rm E}^{\rm s,s}_{\rm sp}+ {\rm E}^{\rm s,p}_{\rm sp} + {\rm E}^{\rm p,s}_{\rm sp} + {\rm E}^{\rm p,p}_{\rm sp} $,
where the superscripts denote which surface potentials contribute to each term. For example,
let us consider the ${\rm E}^{\rm s,s}_{\rm sp}$-term

\begin{widetext}

\begin{eqnarray}
&&{\rm E}^{\rm s,s}_{\rm sp} =  \frac{\varepsilon_o}{2} R \sinh \Lambda \sum_{\ell=0}^\infty \sum_{k = -\ell}^\ell \frac{\lambda_\ell}{2\pi}(-1)^k \int d\Omega \int d\Omega'  
e^{i k (\phi-\phi')} P^k_\ell(\cos \xi) P^{-k}_\ell(\cos \xi')   \nonumber \\
&&
\times 
\frac{V_{\rm s}(\Omega) V_{\rm s}(\Omega')}{ \sqrt{\cosh \Lambda -\cos \xi} \sqrt{\cosh \Lambda -\cos \xi'} }
\left( 
\lambda_\ell \coth \lambda_\ell \Lambda + \frac{\sinh L}{2(\cosh \Lambda - \cos \xi)} 
\right) .
\label{Ess}
\end{eqnarray}

In order to simplify the above expression we first notice that 
the second term in the parenthesis is independent of $\ell$, which allows us to use the
completeness relation for the associated Legendre polynomials 
\begin{eqnarray}
\sum_{\ell=0}^\infty \sum_{k = -\ell}^\ell \frac{\lambda_\ell}{2\pi}(-1)^k 
e^{i k (\phi-\phi')} P^k_\ell(\cos \xi) P^{-k}_\ell(\cos \xi')   
= \delta(\phi-\phi') \delta(\cos \xi - \cos \xi') ,
\label{completeness-rel}
\end{eqnarray}
and perform the $\Omega$ integral. We also notice that the first term in 
the parenthesis can be written as
$\coth \lambda_\ell \Lambda = 2 {\sum_{n=0}^\infty}'  e^{-2 \lambda_\ell \Lambda n}$, 
where the prime in the summation symbol indicates that the $n=0$ term has to
be taken with half weight.
By using the completeness relation Eq.(\ref{completeness-rel}), the addition theorem for Legendre
polynomials, and the identity Eq.(\ref{identity-1}) we can express ${\rm E}^{\rm s,s}_{\rm sp} $ as 
\begin{eqnarray}
&&{\rm E}^{\rm s,s}_{\rm sp} =  \frac{\varepsilon_o}{4} R  \int d\Omega 
\frac{ \sinh^2 \Lambda \ V^2_{\rm s}(\Omega)}{(\cosh \Lambda -\cos \xi)^2} 
+
\frac{\varepsilon_o}{2 \pi}  {\sum_{n=0}^\infty}' e^{-\Lambda n} \int d\Omega \int d\Omega'  
\frac{ a \ V_{\rm s}(\Omega) V_{\rm s}(\Omega') \ \mathcal{G}( e^{- 2 \Lambda n},\cos \gamma ) }{ \sqrt{\cosh \Lambda -\cos \xi} \sqrt{\cosh \Lambda -\cos \xi'} } ,
\end{eqnarray}
where we have introduced the function $\mathcal{G}(t,u)$, defined in Eq.(\ref{identity-1}).
Using the identities in the previous Appendix, that relate bispherical and spherical 
coordinates, we obtain
\begin{eqnarray}
&&{\rm E}^{\rm s,s}_{\rm sp} =  \frac{\varepsilon_o}{4} R  \int d\Omega_{\rm s} 
V^2_{\rm s}(\Omega_{\rm s})
+
\frac{\varepsilon_o}{2 \pi}  {\sum_{n=0}^\infty}' e^{-\Lambda n} \int d\Omega_{\rm s} \int d\Omega_{\rm s'}  
\frac{ R \sinh^3 \Lambda \ V_{\rm s}(\Omega_{\rm s}) V_{\rm s}(\Omega_{\rm s}') \ 
\mathcal{G}( e^{- 2 \Lambda n},\cos \gamma^{\rm s,s}) }{(\cosh \Lambda + \cos \theta)^{3/2} (\cosh \Lambda + \cos \theta')^{3/2} } ,
\end{eqnarray}
where $\cos \gamma^{\rm s,s}$ is given by (\ref{cosgamma}) for the two bispherical coordinates
$(\xi,\phi)$ and $(\xi',\phi')$ written in terms of spherical coordinates, namely
\begin{equation}
\label{ }
\cos \gamma^{\rm s,s} = \frac{ (1 + \cosh \Lambda \cos\theta)(1 + \cosh \Lambda  \cos \theta') + \sinh^2 \Lambda \sin \theta \sin \theta' \cos(\phi-\phi')}{(\cosh \Lambda + \cos\theta)(\cosh \Lambda + \cos \theta') }.
\end{equation}

The same kind of calculations can be performed for the other terms 
${\rm E}^{\rm s,p}_{\rm sp}$, ${\rm E}^{\rm p,s}_{\rm sp}$, and
${\rm E}^{\rm p,p}_{\rm sp}$ in the energy (\ref{exact-energy}).
In this way we derive the energy kernels that appear in (\ref{energy-xfrm}):
\begin{align}
{\mathcal E}_{\rm s,s}(\Omega_{\rm s}; \Omega_{\rm s}' ;D)   = & 
\underbrace{ \frac{\varepsilon_o R}{4}
\delta(\phi-\phi') \delta(\cos\theta-\cos\theta')
+
\frac{\varepsilon_o}{4 \pi}  
R  \lim_{\Lambda\to \infty} \mathcal{G}(1, \cos \gamma^{\rm s,s})  }_{  {\rm self-energy} }  \nonumber \\
& +
\frac{\varepsilon_o}{2 \pi} {\sum_{n=1}^\infty} \frac{ R \sinh^3 \Lambda \ e^{ - \Lambda n} \ 
\mathcal{G}(e^{-2 \Lambda n}, \cos \gamma^{\rm s,s})  }{(\cosh \Lambda + \cos \theta)^{3/2} (\cosh \Lambda + \cos \theta')^{3/2}}  , 
\label{Ess} 
\\
{\mathcal E}_{\rm s,p}(\Omega_{\rm s}; \Omega_{\rm p} ;D)  = &
{\mathcal E}_{\rm s,p}(\Omega_{\rm s}; \Omega_{\rm p} ;D)  =
-\frac{\sqrt{2} \varepsilon_o}{ \pi} \sum_{n=0}^\infty 
\frac{ R^2 \sinh^2 \Lambda \ e^{ - \lambda_n \Lambda} \  \mathcal{G}(e^{-2 \lambda_n \Lambda }, 
\cos \gamma^{\rm s,p})
}{(\cosh \Lambda + \cos \theta)^{3/2} (R^2 \sinh^2 \Lambda + \rho^2)^{3/2}} , 
 \\
{\mathcal E}_{\rm p,p}(\Omega_{\rm p}; \Omega_{\rm p}' ;D) =& 
\underbrace{- \frac{ \varepsilon_0}{4 \pi} 
\frac{ 1}{| {\bf x} - {\bf x}' |^3 } }_{{\rm self-energy} }  + \frac{4 \varepsilon_0}{ \pi}  
{\sum_{n=1}^\infty} \frac{ R^3 \sinh^3 \Lambda \ e^{ - \Lambda n} \   
\mathcal{G}(e^{-2 \Lambda n}, \cos \gamma^{\rm p,p})}{(R^2 \sinh^2 \Lambda + \rho^2)^{3/2} (R^2 \sinh^2 \Lambda + {\rho'}^2)^{3/2}}  . \label{Epp}
\end{align}
The dependency of these kernels on the sphere-plane separation $D$ is encoded
in $\Lambda $, which we recall depends on $D$ as $\cosh \Lambda = 1+D/R$. The functions
$\cos \gamma^{\rm s,p}$ and $\cos \gamma^{\rm p,p}$ are obtained by expressing
(\ref{cosgamma}) in terms of the corresponding spherical or polar coordinates:
\begin{eqnarray}
\cos \gamma^{\rm s,p} &=& \frac{\rho^2 - R^2 \sinh^2 \Lambda}{\rho^2 + R^2 \sinh^2 \Lambda}  \frac{1+ \cosh \Lambda \cos \theta}{\cosh \Lambda + \cos \theta}+  \frac{2 R \rho \sinh \Lambda }{\rho^2 + R^2 \sinh^2 \Lambda}  \frac{  \sinh \Lambda \sin \theta}{\cosh \Lambda + \cos \theta} \cos(\phi-\phi') ,\\
\cos \gamma^{\rm p,p} &=& \frac{(\rho^2 - R^2 \sinh^2 \Lambda) ( {\rho'}^2- R^2 \sinh^2 \Lambda) + 4 R^2 \rho \rho'  \sinh^2 \Lambda  \cos(\phi-\phi') }{(\rho^2 + R^2 \sinh^2 \Lambda)({\rho'}^2 + R^2 \sinh^2 \Lambda)}.
\end{eqnarray}
The first two terms in (\ref{Ess}) and the first term in (\ref{Epp}) correspond to the self-energies of the sphere and the plane, respectively, already derived
in the previous Appendix.  There should be removed from the energy kernels when computing the interaction energy or force.

\end{widetext}


\section{Equipotential case}

In this Appendix we show how to obtain from our general formula (\ref{exact-energy})
the electrostatic energy when both the sphere and the plane are equipotentials.
First we will consider the component of the energy depending quadratically on the sphere's potential
\begin{widetext}

\begin{eqnarray}
{\rm E}^{\rm s,s}_{\rm eq-p} &= & \frac{\varepsilon_o}{2} V^2_{\rm s} \sum_{\ell=0}^\infty \sum_{k = -\ell}^\ell \frac{\lambda_\ell}{2\pi}(-1)^k \int d\Omega \int d\Omega'  e^{-i k (\phi-\phi')} P^k_\ell(\cos \xi) P^{-k}_\ell(\cos \xi')   \nonumber \\
&& \times \bigg\{ 
\frac{a  }{\sqrt{\cosh \Lambda -\cos \xi}}  \bigg[
\frac{1}{\sqrt{ \cosh \Lambda - \cos \xi'}} \bigg( 
\lambda_\ell \coth \lambda_\ell \Lambda + \frac{\sinh \Lambda}{2(\cosh \Lambda - \cos \xi)} \bigg)  \bigg] .
\label{equipotential}
\end{eqnarray}

\end{widetext}
The $\phi$-integrals can be done directly giving Kronecker deltas which collapse the sum on $k$, and the integrals over $\xi$ can be done after using the identity $\frac{1}{\sqrt{\cosh \Lambda - x}} = \sqrt{2} e^{-\Lambda/2} \sum_{n=0}^\infty P_n(x) e^{-\Lambda n}$,
and the orthonormality properties of the Legendre polynomials giving the following identities
\begin{eqnarray}
\label{identity-1}
&& \int_0^\pi d\xi \sin \xi P_\ell(\cos \xi) \frac{1}{\sqrt{ \cosh \Lambda - \cos \xi}} = \frac{ \sqrt{2}}{\lambda_\ell} e^{ - \lambda_\ell \Lambda} , \nonumber \\
&& \int_0^\pi d\xi \sin \xi P_\ell(\cos \xi) \frac{1}{( \cosh \Lambda - \cos \xi)^{3/2}} = \frac{ 2 \sqrt{2}}{\sinh \Lambda} e^{ - \lambda_\ell \Lambda}. \nonumber
\end{eqnarray}
After performing these integrations Eq.(\ref{equipotential}) reduces to
\begin{align}
{\rm E}^{\rm s,s}_{\rm eq-p} = & 2 \pi \varepsilon_o a V^2_{\rm s} \sum_{\ell=0}^\infty   
e^{ - 2 \lambda_\ell \Lambda} ( 
 \coth \lambda_\ell \Lambda + 1 ) \nonumber \\
  =  & 2 \pi \varepsilon_o a V^2_{\rm s} \sum_{\ell=0}^\infty   
\frac{e^{ -  \lambda_\ell \Lambda}}{  
 \sinh \lambda_\ell \Lambda }.
\label{}
\end{align}
Note that the previous equation can be written as 
\begin{align}
{\rm E}^{\rm s,s}_{\rm eq-p}  
  =  & 2 \pi \varepsilon_o a V^2_{\rm s} \sum_{\ell=1}^\infty   
\bigg[ \frac{e^{ -  \ell \frac{\Lambda}{2} }}{  
 \sinh \ell \frac{\Lambda}{2} }- \frac{e^{ - \ell \Lambda}}{  
 \sinh \ell \Lambda } \bigg] ,
\label{}
\end{align}
which after some rearrangement reduces to the well-known form for the energy for the prescribed case of equipotentials  \cite{Smythe} 
\begin{equation}
\label{smythe-energy}
{\rm E}^{s,s}_{\rm eq-p} = 2 \pi \varepsilon_o V_{\rm s}^2  a  \sum_{n= 1}^\infty 
\frac{1}{ \sinh \Lambda n} . 
\end{equation}
With the same identities one can derive the other terms contributing to the electrostatic energy. The cross terms are given by
\begin{align}
{\rm E}^{\rm s,p}_{\rm eq-p} =  \rm{E}^{\rm p,s}_{\rm eq-p} = & -   
 2 \pi \varepsilon_o a V_{\rm s} V_{\rm p} \sum_{\ell=0}^\infty   
\frac{e^{ -  \lambda_\ell \Lambda}}{  
 \sinh \lambda_\ell \Lambda },
\label{}
\end{align}
and the component depending on the square of the plane's potential can be written as 

\begin{align}
\rm{E}^{\rm p,p}_{\rm eq-p} = &    
 2 \pi \varepsilon_o a  V^2_{\rm p} \sum_{\ell=0}^\infty   
  \coth \lambda_\ell \Lambda .
\label{}
\end{align}
The contribution to the electrostatic energy from the preceding equation contains a divergence which results from the infinite self-energy of the plane. Note from Eq.(\ref{self-energy-plane}) and using Eq.(\ref{identity-1}) that for the case of an equipotential on the plane the plane self-energy reduces to
${\rm E}^{\rm p,p}_{\rm eq-p,self-energy} =    2 \pi \varepsilon_o a  V^2_{\rm p} \sum_{\ell=0}^\infty   1$.
By subtracting this divergence the plane-plane contribution to the electrostatic energy becomes

\begin{align}
\rm{E}^{\rm p,p}_{\rm eq-p} -  \rm{E}^{\rm p,p}_{\rm eq-p,self-energy} = &    
 2 \pi \varepsilon_o a  V^2_{\rm p} \sum_{\ell=0}^\infty   
 \frac{2}{e^{ 2\lambda_\ell \Lambda} - 1}
 \nonumber \\
 = & 2 \pi \varepsilon_o a  V^2_{\rm p} \sum_{\ell=0}^\infty   
\frac{e^{ -  \lambda_\ell \Lambda}}{  
 \sinh \lambda_\ell \Lambda }.
\label{}
\end{align}
 By combining all of the terms which contribute to the energy after all divergences have been removed we find the final form for the equipotential energy
 
 \begin{align}
\label{}
\label{}
{\rm E}_{\rm eq-p} = 2 \pi \varepsilon_o (V_{\rm s} - V_{\rm p})^2  a  \sum_{n= 1}^\infty 
\frac{1}{ \sinh \Lambda n} . 
\end{align}

The expression for the force in the equipotential case can be derived by taking a
derivative with respect to sphere-plane separation giving 
\begin{equation}
\label{eq-p}
{\rm F}_{\rm eq-p} =  - 2 \pi \varepsilon_o (V_{\rm s} - V_{\rm p})^2   \sum_{n= 1}^\infty \frac{\coth \Lambda - n \coth \Lambda n}{\sinh \Lambda  n} ,
\end{equation}
which, as expected, agrees with the known results \cite{Smythe}.


\newcommand{\Review}[1]{{\em #1}}
\newcommand{\Volume}[1]{\textbf{#1}}
\newcommand{\Book}[1]{\textit{#1}}
\newcommand{\Eprint}[1]{\textsf{#1}}
\def\etal{\textit{et al}}


\begin{references}
\bibitem{explanation}
To understand this point consider a single patch on the plane with constant potential.
Let us assume that its central point is described by $x \hat{x}$. In bispherical coordinates this point is equivalent to $ \frac{a \sin \xi \cos \phi}{1 - \cos \xi} \hat{x} $.
In polar coordinates the central position of this patch is independent of the 
sphere-plane distance. Alternatively, the patch's position described by bispherical coordinates changes as we alter 
the sphere-plane separation (note the dependence of $a$ on $D$). Therefore, if one were to compute the force by taking the derivative of Eq.(\ref{exact-energy}) derivatives of the potentials on each object would appear in the final expression.
\end{references}
\end{document}